\documentstyle{aipproc2}

\def\al{\alpha}
\def\be{\beta}
\def\ga{\gamma}
\def\de{\delta}
\def\ep{\epsilon}

\def\th{\theta}

\def\ka{\kappa}
\def\la{\lambda}

\def\si{\sigma}

\def\ph{\phi}

\def\ps{\psi}

\def\De{\Delta}

\def\cl{{\cal L}}
\def\fr#1#2{{{#1} \over {#2}}}

\def\pt#1{\phantom{#1}}

\def\ket#1{|{#1}\rangle}

\def\half{{\textstyle{1\over 2}}}
\def\frac#1#2{{\textstyle{{#1}\over {#2}}}}

\def\lsim{\mathrel{\rlap{\lower4pt\hbox{\hskip1pt$\sim$}}
    \raise1pt\hbox{$<$}}}
\def\gsim{\mathrel{\rlap{\lower4pt\hbox{\hskip1pt$\sim$}}
    \raise1pt\hbox{$>$}}}
\def\sqr#1#2{{\vcenter{\vbox{\hrule height.#2pt
         \hbox{\vrule width.#2pt height#1pt \kern#1pt
         \vrule width.#2pt}
         \hrule height.#2pt}}}}

\def\lrprtmu{\stackrel{\leftrightarrow}{\partial_\mu}}
\def\lrprtnu{\stackrel{\leftrightarrow}{\partial^\nu}}

\newcommand{\beq}{\begin{equation}}
\newcommand{\eeq}{\end{equation}}
\newcommand{\bea}{\begin{eqnarray}}
\newcommand{\eea}{\end{eqnarray}}
\newcommand{\rf}[1]{(\ref{#1})}

\def\hydrogen{hydrogen}
\def\antihydrogen{antihydrogen}

\begin{document}

\begin{flushright}
{IUHET 395\\}
{August 1998\\}
\end{flushright}

\title{Hydrogen and Antihydrogen Spectroscopy for\\
Studies of CPT and Lorentz Symmetry\footnote
{\footnotesize Presented by V.A.K.
at the 1998 Conference on Trapped Charged Particles 
and Fundamental Physics,
Pacific Grove, California, August-September 1998}
\\}

\author{Robert Bluhm$^a$, V.\ Alan Kosteleck\'y$^b$, 
and Neil Russell$^b$}
\address{$^a$Physics Department, Colby College, 
Waterville, ME, 04901 U.S.A.\\
\smallskip
$^b$Physics Department, Indiana University,
Bloomington, IN, 47405 U.S.A.}

\maketitle

\begin{abstract}
A theoretical study of possible signals 
for CPT and Lorentz violation 
arising in hydrogen and antihydrogen spectroscopy
is described.
The analysis uses a CPT- and Lorentz-violating extension
of quantum electrodynamics,
obtained from a general Lorentz-violating extension 
of the minimal standard model
with both CPT-even and CPT-odd terms.
Certain 1S-2S transitions and hyperfine Zeeman lines
exhibit effects at leading order in small CPT-violating couplings.
\end{abstract}

\section*{INTRODUCTION}

At presently accessible energy scales,
which are determined by the electroweak scale $m_W$
and are small relative to the Planck mass $M_P$,
the predictions of
the minimal SU(3)$\times$SU(2)$\times$U(1) standard model
appear to be in agreement with nature. 
However,
at scales closer to $M_P$
this model is expected to be superseded by
a fundamental theory that also combines 
quantum mechanics and gravitation in a consistent way.
The fundamental theory is likely to involve
qualitatively new physics as,
for example,
occurs in string (M) theory at the Planck scale.
Associated low-energy signals may exist.
However,
approximately 17 orders of magnitude separate
$m_W$ from $M_P$,
so effects specific to the fundamental theory 
and accessible via existing techniques 
are likely to be heavily suppressed.
Experiments that search for effects forbidden
in the usual renormalizable gauge theories
and that are of high precision
are therefore of particular interest.

In this talk,
the idea is considered that the new physics includes 
a spontaneous violation of Lorentz symmetry
\cite{kps}.
If a theory with Lorentz-covariant dynamics
involves Lorentz-tensor interactions acting 
to destabilize the naive vacuum,
some finite Lorentz-tensor expectation values may arise. 
This can occur in some string theories, for instance.
In the low-energy theory
at the level of the standard model,
apparent Lorentz violations would ensue 
if the orientation of the tensor expectation values
includes the physical four spacetime dimensions.

The CPT theorem connects Lorentz transformations
to the discrete charge-conjugation (C),
parity-reflection (P),
and time-reversal (T) transformations
\cite{cpt}.
It implies that all
local relativistic quantum field theories
satisfying mild technical assumptions
are invariant under CPT.
This suggests that both CPT and Lorentz violations 
represent unconventional and potentially observable effects 
emerging from a fundamental theory.
However,
the heavy suppression expected from the hierarchy 
between $m_W$ and $M_P$
implies that detection of these effects
would be feasible only in 
particularly sensitive experiments.

\section*{EXTENDED QUANTUM ELECTRODYNAMICS}

At the level of the minimal standard model,
the consequences of spontaneous Lorentz and CPT breaking  
can be investigated by incorporating possible terms 
that would represent violations of these symmetries.
There exists a general Lorentz-violating extension 
of the minimal SU(3) $\times$ SU(2)$ \times$ U(1) standard model
\cite{ck}.
It includes both CPT-even and CPT-odd terms.
To date,
it appears to be the sole existing candidate 
for a consistent standard-model extension 
based on a microscopic description of CPT and Lorentz violation.
In any event,
this theory is necessarily the low-energy limit
of any fundamental theory that contains the standard model 
and incorporates spontaneous CPT and Lorentz violation. 

The standard-model extension is theoretically attractive
for several reasons.
For one,
the usual structure of the gauge invariances and the
spontaneous gauge-symmetry breaking are unaffected,
and energy and momentum are conserved
provided the Lorentz symmetry breaking produces 
position-independent expectation values.
Also, 
standard quantization methods apply,
and the extension is hermitian and power-counting renormalizable.
Even though Lorentz symmetry is spontaneously broken,
various desirable features of Lorentz-covariant theories
such as positivity of the energy and microcausality 
are expected to persist
\cite{ck}.
This is largely a result of the Lorentz covariance of 
the underlying fundamental theory
and the conventional quantum description.
In fact,
invariance under rotations or boosts of the observer's inertial frame
(\it observer \rm Lorentz transformations)
is retained even at the level of 
the standard-model extension.
Only rotations or boosts of particles 
and localized field distributions 
(\it particle \rm Lorentz transformations)
introduce Lorentz breaking,
as a result of couplings to the tensor vacuum expectation values.

Details of the construction and the specific form
of the standard-model extension,
including both CPT-even and CPT-odd terms,
are provided in the literature
\cite{ck}.
Various limits of this theory are 
of direct relevance to experiments testing 
aspects of quantum electrodynamics (QED).
In this talk,
attention is primarily given to the special limit 
that produces a CPT- and Lorentz-violating theory
for a charged fermion interacting via the electromagnetic force
\cite{ck}.
As an explicit example,
here are the terms appearing in the lagrangian
extension of the usual quantum theory of 
photons, electrons, and positrons.
In units with $\hbar = c = 1$,
the standard QED lagrangian is 
\beq
\cl^{\rm QED} =
\overline{\ps} \ga^\mu (\half i \lrprtmu - q A_\mu ) \ps 
- m \overline{\ps} \ps 
- \frac 1 4 F_{\mu\nu}F^{\mu\nu}
\quad .
\label{a}
\eeq
In the fermion sector, there are two CPT-breaking terms:
\beq
\cl^{\rm CPT}_{e} =
- a_{\mu} \overline{\ps} \ga^{\mu} \ps 
- b_{\mu} \overline{\ps} \ga_5 \ga^{\mu} \ps \quad ,
\quad ,
\label{b}
\eeq
while there is one possibility in the photon sector:
\beq
\cl^{\rm CPT}_{\ga} =
\half (k_{AF})^\ka \ep_{\ka\la\mu\nu} A^\la F^{\mu\nu}
\quad .
\label{c}
\eeq
The possible Lorentz-violating but CPT-preserving terms
in the fermion sector are:
\beq
\cl^{\rm Lorentz}_{e} = 
c_{\mu\nu} \overline{\ps} \ga^{\mu} 
(\half i \lrprtnu - q A^\nu ) \ps 
+ d_{\mu\nu} \overline{\ps} \ga_5 \ga^\mu 
(\half i \lrprtnu - q A^\nu ) \ps 
- \half H_{\mu\nu} \overline{\ps} \si^{\mu\nu} \ps 
\quad ,
\label{d}
\eeq
while the only possibility in the photon sector is:
\beq
\cl^{\rm Lorentz}_{\ga} =
-\frac 1 4 (k_F)_{\ka\la\mu\nu} F^{\ka\la}F^{\mu\nu}
\quad .
\label{e}
\eeq
In the above expressions,
the unconventional coupling coefficients
govern the magnitude of the CPT- and Lorentz-violating effects
and are expected to depend on the small ratio $m_W/M_P$. 
Note that all the extra couplings are hermitian.
It can be shown 
using field redefinitions
that some coupling-coefficient components
are physically unobservable.
The reader is referred to the literature \cite{ck}
for details about this and other issues,
and for more information about the notation used above.

\section*{OVERVIEW OF SOME EXPERIMENTAL TESTS}

Most experiments testing Lorentz invariance or CPT symmetry
are likely to be insensitive to the extra couplings 
in the standard-model extension 
due to the expected heavy suppression factors. 
A few experiments of exceptional sensitivity
could bound or in principle detect these effects
despite the suppression.
In such cases,
the standard-model extension can be used as a 
quantitative theoretical guide to potential experimental signals.
It also offers the possibility of
analyzing and comparing bounds on CPT and Lorentz violation
arising from different experiments.

At present,
implications of the standard-model extension 
have been studied for CPT and Lorentz tests that involve: 
observations of neutral-meson oscillations
\cite{kexpt,ckpv,bexpt,ak},
measurements of particle and antiparticle properties
in Penning traps
\cite{pennexpts,bkr},
spectroscopic comparisons of  
hydrogen and antihydrogen
\cite{ce,bkr2},
determination of photon properties
\cite{ck},
and baryon-number generation
\cite{bckp}.
A variety of additional studies are in progress,
notably one \cite{kla}
establishing the implications for the
standard-model extension of
high-precision clock-comparison experiments
\cite{cc}.

This section of the talk provides a short summary
of a subset of the results obtained.
The use of hydrogen and antihydrogen spectroscopy
to test CPT and Lorentz symmetries
is described in the following three sections.

The flavor oscillations of certain neutral-meson systems
provide a valuable interferometric tool for studying
CP violation.
The effective hamiltonian for the
time evolution of a $P$-meson state,
where $P$ represents one of 
the neutral $K$, $D$, $B_d$, or $B_s$ mesons,
depends on two kinds of (indirect) CP violation.
The first involves T violation with CPT invariance
and is conventionally described with a 
complex parameter $\ep_P$.
The second involves CPT violation with T invariance
and is described with
a complex parameter $\de_P$.
The standard-model extension can be used to
derive an expression for $\de_P$
\cite{ak}.

It turns out that
flavor oscillations in neutral-$P$ systems
are sensitive to only one type of CPT-violating  
term in the standard-model extension,
$- a^q_{\mu} \overline{q} \ga^\mu q$,
where $q$ is a quark field
and $a^q_{\mu}$ is a spacetime-constant coupling coefficent
with value dependent on the quark flavor $q$.
None of the other experiments discussed in this talk 
involve flavor changes,
and it has been shown that
as a result these other experiments
are insensitive to $a^q_{\mu}$-type coupling coefficients.
In this respect,
the bounds on CPT violation from neutral-meson tests of CPT
are entirely disjoint from those of other experiments.

In the observer frame in which the 
Lorentz-violating coupling coefficients are defined,
denote the $P$-meson four-velocity by
$\be^\mu \equiv \ga(1,\vec\be)$.
Then,
at leading order in all the standard-model coupling coefficients,
the expression for $\de_P$ is \cite{ak}
\beq
\de_P \approx 
\fr{\ga(\De a_0 - \vec \be \cdot \De \vec a)}{\De m} ~ i 
\sin\hat\ph e^{i\hat\ph} 
\quad . 
\label{f}
\eeq
In this equation,
$\De a_\mu \equiv a_\mu^{q_2} - a_\mu^{q_1}$,
where $q_1$ and $q_2$ represent the valence-quark flavors 
in the $P$ meson.
Also,
$\hat\ph\equiv \tan^{-1}(2\De m/\De\ga)$,
where the mass and decay-rate differences
between the $P$-meson eigenstates are,
respectively,
$\De m$ and $\De \ga$.

The expression for $\de_P$
implies a proportionality between 
the real and imaginary parts of $\de_P$
\cite{ckpv}.
Note that
the magnitude of $\de_P$ can vary with $P$ 
because the couplings $a_\mu^q$ are flavor dependent
\cite{ckpv},
so the magnitude of CPT-violating effects 
may differ in distinct neutral-meson systems.
For instance,
the magnitude of CPT violation might
grow with the mass of the quarks involved,
as the Yukawa couplings do in the standard model.
Also, 
the explicit dependence in Eq.\ \rf{f} of $\de_P$ on
the boost magnitude and orientation
implies several types of potentially observable effect
including,
for instance,
larger CPT-violating effects in boosted mesons
\cite{ak}.
Experiments involving mesons with different momenta
may therefore have different CPT reaches.
The best reported bounds to date
come from the kaon system
\cite{kexpt}.
Recently, two CERN experiments \cite{bexpt}
have obtained results for the $B_d$ system,
following the observation \cite{ckpv}
that existing data already suffice to yield CPT limits.
Other studies are ongoing.

A number of experiments 
that test CPT and Lorentz symmetries
in a different way 
have been performed with the goal of comparing 
particle and antiparticle properties.
An important technique is the use of a Penning trap
to confine single particles over relatively large
time scales while high-precision measurements
are taken of properties such as
anomaly and cyclotron frequencies
\cite{pennexpts}.
Experiments of this type can constrain,
for example,
the coupling coefficients in the
fermion sector of the extended QED.
Possible observable signals in the context of this theory,
the corresponding relevant figures of merit,
and the associated CPT and Lorentz reaches
have been obtained
\cite{bkr}.
As just one example,
using existing technology 
and implementing a relatively minor change in 
experimental procedure,
Penning-trap experiments comparing the anomalous magnetic moments
of electrons and positrons could place
a bound of roughly $10^{-20}$
on a figure of merit involving 
the spatial components of the coefficient $b_\mu$.

The extra terms \rf{c} and \rf{e} in the QED extension
represent modifications to photon properties.
It turns out that the ensuing generalized Maxwell equations 
describe two independent propagating degrees of freedom
as in the conventional case
\cite{ck}.
Typically, however,
each has a distinct dispersion relation,
which implies several interesting effects.
For example, 
the vacuum becomes birefringent,
so that in the presence of the CPT and Lorentz violation 
an electromagnetic wave propagating in the vacuum
exhibits properties similar to those displayed 
by conventional radiation traveling
in an optically anisotropic and gyrotropic transparent crystal 
having spatial dispersion of the axes.
Behavior of this type can be constrained
from the observed absence of birefringence 
on radio waves propagating over cosmological distances. 
The components of the CPT-odd coefficient $(k_{AF})_\mu$ 
are presently bounded to $\lsim 10^{-42}$ GeV
\cite{cfj,hpk},
although a disputed claim \cite{nr,misc}
exists for a nonzero effect with 
$|\vec k_{AF}|\sim 10^{-41}$ GeV.
The rotation-invariant irreducible component of 
the CPT-even coefficient $(k_F)_{\ka\la\mu\nu}$ 
is bounded to $\lsim 10^{-23}$ 
by cosmic-ray existence \cite{cg}
and other experiments.
The rotation-violating irreducible components
of $(k_F)_{\ka\la\mu\nu}$ 
could in principle be bounded to about $10^{-27}$
with existing techniques
seeking cosmological birefringence
\cite{ck},
but no actual limit has been obtained to date.

The CPT-even term in Eq.\ \rf{e}
introduces no theoretical difficulties.
However,
the CPT-odd term in Eq.\ \rf{c} can generate 
negative contributions to the energy
\cite{cfj}.
This may represent a theoretical difficulty
and indicates $(k_{AF})^\ka$ vanishes
\cite{ck},
which would be in agreement with the
tight experimental bound from cosmological birefringence.
It can be argued that a zero value of $(k_{AF})_\mu$ 
is acceptable theoretically despite the
possibility of radiative corrections 
from diagrams involving the CPT-violating couplings
in the fermion sector
because the one-loop effects are finite.

\section*{1S-2S SPECTROSCOPY IN FREE HYDROGEN AND ANTIHYDROGEN}

The remainder of this talk
addresses the possibility of searching for CPT and Lorentz violations
by making high-precision comparisons 
of the spectra of \hydrogen\ and \antihydrogen\
\cite{bkr2}.
The feasibility of the idea of comparative tests 
\cite{ce}
has received a boost
following the recent production and observation of antihydrogen 
\cite{oelert,mandel},
and several proposals for \antihydrogen\ spectroscopy
have been advanced.
In the near future,
the \antihydrogen\ fine structure 
and Lamb shift may be obtained 
within a few percent
by observations on a relativistic \antihydrogen\ beam 
\cite{mandel2}.
A more ambitious goal is to measure 
the two-photon 1S-2S transition
in \antihydrogen, 
which is expected to have a natural linewidth
of only $1.3$ Hz and is therefore 
a promising candidate for high-precision spectroscopy.
Proposed experiments 
\cite{gab2}
would provide a comparison of the 1S-2S transitions 
in spin-polarized \hydrogen\ and \antihydrogen\ 
confined within a magnetic trap.
For \hydrogen, 
a cold atomic beam 
has been used to measure the 1S-2S transition frequency 
to $3.4$ parts in $10^{14}$
\cite{hansch},
while trapping techniques have yielded 
a frequency precision of about $10^{-12}$
\cite{cesar}.
A limiting accuracy of about $10^{-18}$ may be attainable
\cite{hanschICAP}.

A theoretical analysis of signals for CPT and Lorentz violations
in hydrogen and antihydrogen spectroscopy 
is feasible 
\cite{bkr2}
in the context of the
QED extension described in the second part of this talk.
In this section,
possible effects on the free-atom 1S-2S transition
are considered.
These are relevant,
for example,
to the experiments with cold atomic beams of hydrogen
\cite{hansch}.
The next section treats the trapped-atom case. 
A detailed theoretical treatment of the proposed
experiments with relativistic beams 
\cite{mandel2},
which are expected to have significantly poorer
frequency resolutions than those based on other techniques,
remains to be performed and is not discussed here.
Note, however,
that all the experimental situations discussed below
are sensitive only to spatial or mixed spatio-temporal components
of the CPT- and Lorentz-violating couplings
in the comoving Earth frame,
whereas a boost can induce sensitivity to purely timelike components
and can enhance CPT- and Lorentz-violating effects
\cite{ak}.

To calculate effects on the free-atom 1S and 2S energy levels,
the modified Dirac equation for 
a four-component electron field $\ps$ 
in the proton Coulomb potential 
\beq
A^\mu = \fr{|e|}{4 \pi r}  ( 1, \vec 0 )
\quad 
\label{defsa}
\eeq
is needed.
The desired equation is found from
Eqs.\ \rf{a}, \rf{b}, and \rf{d} to be
\beq
\left( i \ga^\mu D_\mu - m_e - a_\mu^e \ga^\mu
- b_\mu^e \ga_5 \ga^\mu 
- \half H_{\mu \nu}^e \si^{\mu \nu} 
+ i c_{\mu \nu}^e \ga^\mu D^\nu 
+ i d_{\mu \nu}^e \ga_5 \ga^\mu D^\nu \right) \ps = 0
\quad ,
\label{dirac}
\eeq
where 
$m_e$ is the electron mass
and the covariant derivative is
\beq
i D_\mu \equiv i \partial_\mu - q A_\mu
\quad 
\label{deriv}
\eeq
with the electron charge being $q = -|e|$.
Both a free electron and a free proton
have distinct CPT- and Lorentz-violating coupling coefficients
in the typical case \cite{ck,bkr},
so superscripts $e$ have been added to the couplings
in Eq.\ \rf{dirac}.
In what follows,
the corresponding couplings for a free proton
are denoted by
$a_\mu^p$, $b_\mu^p$, $H_{\mu \nu}^p$, $c_{\mu \nu}^p$, 
$d_{\mu \nu}^p$.
Note that,
as mentioned following Eq.\ \rf{e},
certain combinations of the electron and proton couplings
can be shown on general grounds to be physically unobservable
\cite{ck}.
This is true,
for example,
of the coefficients $a_\mu^e$ and $a_\mu^p$.
Although all couplings are kept explicitly 
in the derivations that follow,
it is to be expected that 
the ensuing possible spectroscopic signals
in hydrogen and antihydrogen
are independent of the unobservable couplings. 

Since the coupling coefficients are expected
to be highly suppressed,
it is reasonable to calculate
the dominant effects on the hydrogen and antihydrogen spectra
via perturbation theory in relativistic quantum mechanics.
The relevant unperturbed hamiltonians
and the corresponding eigenstates 
are identical 
for hydrogen and antihydrogen,
as are all perturbative effects
from conventional quantum electrodynamics.
The unconventional coupling coefficients
introduce hermitian perturbations
that can differ for hydrogen and antihydrogen.
For the electron and positron,
the explicit forms of these perturbations
follow from Eq.\ \rf{dirac}
after application of suitable field redefinitions
to obtain the hamiltonian and,
for the positron,
a standard charge-conjugation procedure
\cite{bkr}.
The CPT and Lorentz violations from the proton sector
also produce energy perturbations,
which at leading order can be derived 
using relativistic two-fermion techniques
\cite{D2}.

In what follows,
the uncoupled angular-momentum quantum numbers
for the S-state electron/positron and for the proton/antiproton 
are denoted by $J=1/2$ and $I=1/2$, respectively.
Their components along the spin-quantization axis are
$m_J$, $m_I$,
and the corresponding basis states are denoted $\ket{m_J,m_I}$.
Note that distinct real experiments are likely to involve
different spin-quantization axes
relative to any single specified inertial frame,
so comparisons between various experiments may require care 
in allowing for possible geometrical factors.

The result of the perturbative calculation
is that the 1S and 2S levels in hydrogen 
are shifted by identical amounts
$\De E^{H}$
\cite{bkr2}:
\bea
\De E^{H} (m_J, m_I)
& \approx &
(a_0^e + a_0^p - c_{00}^e m_e - c_{00}^p m_p)
+ (-b_3^e + d_{30}^e m_e + H_{12}^e) 
\fr {m_J}{|m_J|} 
\nonumber\\
&&
\pt{(a_0^e + a_0^p - c_{00}^e m_e - c_{00}^p m_p)}
+ (-b_3^p + d_{30}^p m_p + H_{12}^p) 
\fr {m_I}{|m_I|}
\quad ,
\label{EHJI}
\eea
where $m_p$ is the proton mass.

A similar calculation for antihydrogen also yields
equal 1S and 2S level shifts
$\De E^{ \overline{H}}$,
given by Eq.\ \rf{EHJI}
with the substitutions
\beq
a_\mu^e \rightarrow - a_\mu^e
\quad , \quad
d_{\mu \nu}^e \rightarrow - d_{\mu \nu}^e
\quad , \quad
H_{\mu \nu}^e \rightarrow - H_{\mu \nu}^e
\quad ; \quad
a_\mu^p \rightarrow - a_\mu^p
\quad , \quad
d_{\mu \nu}^p \rightarrow - d_{\mu \nu}^p
\quad , \quad
H_{\mu \nu}^p \rightarrow - H_{\mu \nu}^p
\quad . \quad
\label{subst}
\eeq
Note that in all these expressions
the leading-order contributions from the proton/antiproton
have the same mathematical form as
those from the electron/positron.

The electron (positron) and proton (antiproton)
angular momenta are coupled through the hyperfine interaction.
The relevant basis states are thus
linear combinations $\ket{F,m_F}$
of the $\ket{m_J,m_I}$ states,
where $F$ is the total angular-momentum
quantum number and $m_F$ is its projection
on the quantization axis.
For the two-photon 1S-2S transition,
the selection rules 
are $\De F = 0$ and $\De m_F = 0$
\cite{cagnac},
which allows four 1S-2S transitions 
in \hydrogen\ and four in \antihydrogen.
These transitions involve states with 
identical spin configurations.
However,
for hydrogen the result \rf{EHJI} 
of the perturbative calculation implies
that the leading-order level shifts
for 1S and 2S hydrogen states
with the same spin configuration
are identical.
The same follows from Eq.\ \rf{subst}
for antihydrogen.
Therefore,
the 1S-2S frequencies are unaffected
at leading order for all these transitions.
Indeed,
this result could have been anticipated 
from the discussion in Ref.\ \cite{bkr}
showing that observable CPT-violating effects 
must also involve spin-flip processes and CT violation.

In summary,
\it
no leading-order 1S-2S spectroscopic signal
occurs for Lorentz or CPT violation in 
free \hydrogen\ or in free \antihydrogen
\cite{bkr2}.
\rm

Non-leading level shifts can produce observable signals,
but these are suppressed.
The dominant subleading effects from electron/positron
and proton/antiproton CPT- and Lorentz-violation terms
are relativistic corrections suppressed
by at least $\al^2 \simeq 5\times 10^{-5}$.
As an explicit example,
consider the coupling coefficient $b_\mu^e$
in Eq.\ \rf{dirac}.
If this coupling is nonzero,
the $m_F = 0 \rightarrow m_{F^\prime} = 0$
is unaffected but
a subleading-order frequency shift
in the $m_F = 1 \rightarrow m_{F^\prime} = 1$ transition
appears.
It is given by
\beq
\de \nu^H_{1S-2S} \approx - \fr{\al^2 b_3^e}{8 \pi}
\quad .
\label{shifta}
\eeq

The potential signals from subleading effects
are suppressed to the extent that 
feasible $g-2$ experiments
could exclude their observation in free
hydrogen or antihydrogen.
As mentioned above,
an electron-positron $g-2$ comparison
using present technology 
with a minor change in experimental procedure
could attain a tight bound on $b_3^e$ 
\cite{bkr}.
The effect of a nonzero $b_3^e$ at this level 
on the 1S-2S frequency in free hydrogen 
would be to produce a nonzero frequency shift 
$\de \nu^H_{1S-2S} \lsim 5$ $\mu$Hz,
which is below the resolution of the 1S-2S line center.
Similarly,
bounds attainable in Penning-trap experiments 
comparing $g-2$ for protons and antiprotons 
could exclude observable signals in 1S-2S transitions. 
The basic reason why $g-2$ experiments are so effective 
in constraining possible violations
is that they involve spin-flip transitions
that exhibit unsuppressed sensitivity
to CPT and Lorentz breaking.
The $g-2$ experiments have an absolute frequency resolution 
of about 1 Hz.
Although the idealized line-center resolution
for free-hydrogen or free-antihydrogen 1S-2S transitions 
is about three orders of magnitude better, 
the CPT- and Lorentz-violating effects on these transitions 
are suppressed by about five orders of magnitude
and so the net sensitivity of the $g-2$ experiments 
is better.
Note that it is inappropriate in this context to compare
the conventional figure of merit for CPT breaking 
in $g-2$ experiments \cite{pdg},
\beq
r_g = \fr{|g_{e^-} - g_{e^+}|}{g_{\rm av}} \lsim 2 \times 10^{-12}
\quad ,
\label{figmer}
\eeq
with the idealized resolution of the 1S-2S line,
\beq
\De \nu_{1S-2S}/\nu_{1S-2S} \simeq 10^{-18}
\quad .
\label{fm2}
\eeq
These two quantities are physically very different
\cite{bkr}.
A relevant comparison would involve the same physics,
such as the absolute frequency resolution
and sensitivity to CPT- and Lorentz-violating effects used above.

\section*{1S-2S SPECTROSCOPY IN TRAPPED HYDROGEN AND ANTIHYDROGEN}
 
The results in the previous section for free hydrogen
and antihydrogen may be modified in the presence
of external fields,
which can induce transitions
between states with different spin configurations.
External fields are present for the class of proposed experiments
\cite{gab2}
involving spectroscopy of \hydrogen\ or \antihydrogen\ 
confined within a magnetic trap
with an axial bias magnetic field,
such as an Ioffe-Pritchard trap \cite{ip}.
Next,
a theoretical analysis of possible signals
of CPT and Lorentz violation in this context is described.
In what follows,
the four 1S hyperfine Zeeman levels in hydrogen
are denoted by $\ket{a}_1$, $\ket{b}_1$, $\ket{c}_1$, $\ket{d}_1$,
in order of increasing energy in a magnetic field $B$.
The corresponding four 2S levels are denoted
$\ket{a}_2$, $\ket{b}_2$, $\ket{c}_2$, $\ket{d}_2$.
The same notation is used for the 1S and 2S
hyperfine Zeeman levels in antihydrogen.

The Zeeman levels $\ket{a}_n$ and $\ket{c}_n$,
$n= 1,2$,
are mixed-spin states.
For hydrogen,
they are given in terms of the  
basis states $\ket{m_J,m_I}$ by 
\bea
\ket{a}_n &=& \cos \th_n \ket{-\half,\half} -
\sin \th_n \ket{\half,-\half}
\quad , \nonumber \\
\ket{c}_n &=& \sin \th_n \ket{-\half,\half} +
\cos \th_n \ket{\half,-\half}
\quad .
\label{aa}
\eea
The mixing angles $\th_n$ are given by 
\beq
\tan 2 \th_n \approx \fr{(51 {\rm ~mT})}{n^3B}
\quad .
\label{tan2t}
\eeq
Expressions similar to \rf{aa} hold for antihydrogen,
but the spin labels are reversed.

For hydrogen,
in the absence of perturbations and prior to excitation,
the low-field-seeker states $\ket{c}_1$ and $\ket{d}_1$
are confined in the trap.
Spin-exchange collisions 
cause the $\ket{c}_1$ occupation to decrease with time:
$\ket{c}_1 + \ket{c}_1 \rightarrow \ket{b}_1 + \ket{d}_1$.
The primary states in the trap are therefore $\ket{d}_1$.
Moreover,
the transition $\ket{d}_1 \to\ket{d}_2$ is field independent
for small magnetic fields.
It might therefore seem reasonable to perform an experiment
comparing the frequency $\nu^H_d$ 
for the 1S-2S transition $\ket{d}_1 \rightarrow \ket{d}_2$ 
in \hydrogen\ 
with the frequency $\nu^{\overline{H}}_d$ 
for the corresponding transition in \antihydrogen. 
However,
the $\ket{d}_n$ states in hydrogen have no spin mixing,
so the frequency is unaffected to leading order
by CPT- and Lorentz-breaking effects.
A similar result is true for \antihydrogen.
This means that
\cite{bkr2}
\beq
\de \nu^H_d = \de \nu^{\overline{H}}_d \simeq 0
\quad 
\label{resulta}
\eeq
to leading order.

Thus,
\it
no leading-order 1S-2S spectroscopic signal
for Lorentz or CPT violation occurs for unmixed-spin states
in hydrogen or antihydrogen
confined in a magnetic trap with an axial bias field
\cite{bkr2}.
\rm

It therefore appears worthwhile theoretically to examine
1S-2S transitions involving mixed-spin states.
Indeed,
for the $\ket{c}_1 \rightarrow \ket{c}_2$ 
transition in \hydrogen\ 
the spin mixing induces an unsuppressed frequency shift 
\beq
\de \nu_c^H \approx
-\fr{\ka}{2\pi} (b_3^e - b_3^p - d_{30}^e m_e 
+ d_{30}^p m_p - H_{12}^e + H_{12}^p)
\quad ,
\label{nucH}
\eeq
where 
\beq
\ka\equiv \cos 2\th_2 - \cos 2\th_1
\quad .
\label{kappa}
\eeq
The corresponding transition in antihydrogen
in the same magnetic field exhibits a frequency shift
$\de \nu_c^{\overline{H}}$ 
given by an expression of the form \rf{nucH}
except with opposite signs for $b_3^e$ and $b_3^p$.

The unsuppressed sensitivity to CPT and Lorentz breaking
of the $\ket{c}_1 \rightarrow \ket{c}_2$ transition
represents a theoretical advantage of
a factor of about $4/\al^2 \simeq 10^5$
over the suppressed effects from
the $\ket{d}_1 \rightarrow \ket{d}_2$ transition.
Since the frequencies $\nu_c^H$ and $\nu_c^{\overline H}$
vary with the spatial components of the CPT-violating couplings 
$b_\mu^e$ and $b_\mu^p$ in the comoving Earth frame,
they would exhibit diurnal variations.
Moreover,
frequency measurements in a given magnetic trapping field
would also display a nonzero difference
\beq
\De \nu_{1S-2S,c} \equiv \nu_c^H 
- \nu_c^{\overline{H}} \approx - \fr{\ka}{\pi} (b_3^e - b_3^p)
\quad .
\label{nonzero}
\eeq
This difference varies with the amount of spin mixing
according to the parameter $\ka$,
so a maximal value of $\ka$ is theoretically desirable.
It is $\ka \simeq 0.67$,
and it is attained when $B_0 \simeq 0.01$ T.

From the experimental perspective,
the 1S-2S transition
$\ket{c}_1 \rightarrow \ket{c}_2$
is likely to be less advantageous
because it is field dependent  
in both \hydrogen\ and \antihydrogen.
An experiment would therefore have to 
address the issue of Zeeman broadening 
from the inhomogeneous trapping fields.
For instance,
at $B\simeq 10$ mT the 1S-2S linewidth 
for the $\ket{c}_1 \rightarrow \ket{c}_2$
transition is broadened to over 1 MHz 
for both \hydrogen\ and \antihydrogen\
even at a temperature of $100 \mu$K.
Although present methods might reduce the impact of this effect,
it would seem necessary that other techniques be developed
if resolutions of the order of the natural linewidth 
are to be reached.

To summarize,
\it
unsuppressed 1S-2S spectroscopic signals
for Lorentz and CPT violation appear for transitions
involving mixed-spin states in hydrogen or antihydrogen
atoms confined in a magnetic trap with an axial bias field
\cite{bkr2}.
\rm

\section*{HYPERFINE SPECTROSCOPY IN HYDROGEN AND ANTIHYDROGEN}

The remainder of this talk
addresses the issue of possible 
CPT- and Lorentz-violating signals 
in frequency measurements of hyperfine Zeeman transitions
in trapped \hydrogen\ and \antihydrogen
\cite{bkr2}.
The interest in these is partially motivated  
by the resolution below 1 mHz 
that has already been attained 
in transitions between $F = 0$ and $F^\prime = 1$ hyperfine levels
of a hydrogen maser
\cite{ramsey}.

Perturbative calculations along the lines described above
show that all four hyperfine levels 
in the 1S ground state of hydrogen
are shifted by CPT- and Lorentz-violating effects.
One contribution to the shifts,
$a_0^e + a_0^p -c_{00}^e m_e -c_{00}^p m_p$,
is identical for all four levels
and therefore has no effect on any frequencies. 
There are also spin-dependent energy shifts,
given by
\cite{bkr2}
\bea
\De E_a^H &\approx&  
\hat\ka (b_3^e - b_3^p - d_{30}^e m_e 
+ d_{30}^p m_p - H_{12}^e + H_{12}^p)
\quad ,
\nonumber\\
\De E_b^H &\approx& 
b_3^e + b_3^p - d_{30}^e m_e 
- d_{30}^p m_p - H_{12}^e - H_{12}^p
\quad ,
\nonumber\\
\De E_c^H &\approx& -\De E_a^H
\quad ,
\nonumber\\
\De E_d^H &\approx& - \De E_b^H
\quad ,
\label{abcd}
\eea
where 
\beq
\hat\ka \equiv \cos2 \th_1
\quad 
\label{kappahat}
\eeq
is a parameter analogous to $\ka$ of Eq.\ \rf{kappa}
that grows with $B$,
with $\hat\ka \simeq 1$ when $B \simeq 0.3$ T.

To begin,
suppose the magnetic field vanishes.
Then,
$\hat\ka =0$ and so Eq.\ \rf{abcd}
shows that the states $\ket{a}_1$ and $\ket{c}_1$ are unchanged.
However, 
the energies of $\ket{b}_1$ and $\ket{d}_1$ 
shift equally in magnitude but oppositely in sign.
Thus,
even for $B=0$ the three $F=1$ levels are split.

If instead the magnetic field is nonzero,
then the energies of all four hyperfine levels are changed.
Consider first the conventional \hydrogen\ maser,
which uses a small magnetic field 
and involves the (approximately field-independent) transition 
$\ket{c}_1 \rightarrow \ket{a}_1$.
For this situation,
the value of $\hat\ka$ is roughly $10^{-4}$ 
so the spin-mixing is small.
This would act as a suppression factor for 
CPT- and Lorentz-breaking effects
in possible high-precision measurements of the maser $\si$ line 
$\ket{c}_1 \rightarrow \ket{a}_1$.

In contrast,
unsuppressed frequency differences appear
between the field-dependent transitions 
$\ket{d}_1 \rightarrow \ket{a}_1$
and $\ket{b}_1 \rightarrow \ket{a}_1$.
Equation \rf{abcd} gives
\beq
|\De \nu_{d-b}^H| \approx
\fr 1 {\pi} |b_3^e + b_3^p - d_{30}^e m_e - d_{30}^p m_p
- H_{12}^e - H_{12}^p|
\quad .
\label{freqdiff}
\eeq
This difference would vary diurnally 
in the comoving Earth frame,
as occurs with the shifts \rf{nucH},
so in principle a measurement of 
$|\De \nu_{d-b}^H|$ 
in hydrogen alone could provide
a signal of CPT and Lorentz violation.
However,
in practice the attainable frequency resolution
is likely to be affected by the broadening 
due to field inhomogeneities.
An experiment of this type would also need to address the issue
of distinguishing the signal from 
possible backgrounds due to residual Zeeman splittings.

Instead,
one could envisage
using a field-independent transition point
to minimize the frequency dependence on the magnetic field
and making a direct comparison of hydrogen and
antihydrogen transition frequencies
to avoid issues with the background splittings.
Consider,
for example,
an experiment performing 
high-resolution radiofrequency spectroscopy
in trapped \hydrogen\ and \antihydrogen\
on the $\ket{d}_1 \rightarrow \ket{c}_1$ transition
at the field-independent transition point $B \simeq 0.65$ T.
To avoid Doppler broadening,
cooling to temperatures of 100 $\mu$K
with a good signal-to-noise ratio 
is likely to be needed.
Also,
the relatively high bias field suggests 
potentially larger field inhomogeneities would occur,
so a stiff box shape would be preferable for the trapping potential.
Under these circumstances,
it may be possible to attain frequency resolutions of order 1 mHz.

In a magnetic field of 0.65 T,
the state $\ket{c}_1$ in hydrogen 
is well approximated as a spin-polarized
level with $\ket{m_J, m_I}=\ket{1/2,-1/2}$ .
This means that the transition of interest,
$\ket{d}_1 \rightarrow \ket{c}_1$,
involves a proton spin flip,
which in turn implies a signal dependence
only on CPT- and Lorentz-violating effects
for the proton.
Explicit calculation shows that the 
frequency shift for hydrogen is
\beq
\de \nu_{c \rightarrow d}^H \approx
\fr 1 {\pi}
(-b_3^p + d_{30}^p m_p + H_{12}^p)
\quad ,
\label{freqshifth}
\eeq
while that for antihydrogen is
\beq
\de \nu_{c \rightarrow d}^{\overline{H}} \approx 
\fr 1 {\pi}
(b_3^p + d_{30}^p m_p + H_{12}^p)
\quad ,
\label{freqshifthbar}
\eeq
confirming the expected dependence on proton coupling coefficients.

Like the quantity 
$|\De \nu_{d-b}^H|$ of Eq.\ \rf{freqdiff},
diurnal variations of the the frequencies
$\nu_{c \rightarrow d}^H$ and
$\nu_{c \rightarrow d}^{\overline{H}}$ 
would provide a signal for CPT and Lorentz violation.
However,
the difference 
\beq
\De \nu_{c \rightarrow d} \equiv 
\nu_{c \rightarrow d}^H - \nu_{c \rightarrow d}^{\overline{H}}
\approx - \fr{2 b_3^p} {\pi}
\quad 
\label{nudiff}
\eeq
between these frequencies
has the potential to provide an instantaneous,
clean, and accurate test 
of CPT-violating couplings $b_3^p$ for the proton.

Relevant figures of merit for the diurnal
and instantaneous signals in Eqs.\ 
\rf{freqshifth},
\rf{freqshifthbar},
and \rf{nudiff}
(as well as ones for other signals mentioned above
in this and earlier sections)
can be defined following the methods 
developed for Penning-trap tests
\cite{bkr}.
For the instantaneous signal in Eq.\ \rf{nudiff},
an appropriate choice is
\bea
r^H_{rf,c \rightarrow d} & \equiv &
\fr{|({\cal E}_{1,d}^H - {\cal E}_{1,c}^H)
- ({\cal E}_{1,d}^{\overline{H}} - {\cal E}_{1,c}^{\overline{H}})|}
{{\cal E}_{1,{\rm av}}^H}
\nonumber \\
&\approx &
\fr{2\pi |\De \nu_{c \rightarrow d}|}{m_H}
\quad ,
\label{rrf}
\eea
where $m_H$ is the atomic mass of hydrogen
and where 
the relativistic energies in the ground-state hyperfine levels
are denoted by ${\cal E}_{1,d}^H$, ${\cal E}_{1,c}^H$ 
for hydrogen and by 
${\cal E}_{1,d}^{\overline{H}}$, ${\cal E}_{1,c}^{\overline{H}}$
for antihydrogen. 
Suppose,
for instance
that a 1 mHz frequency resolution could indeed be reached
in an experiment of this type.
This would represent an estimated upper bound on
the figure of merit \rf{rrf} of approximately
$r^H_{rf,c \rightarrow d} \lsim 5 \times 10^{-27}$.
The associated constraint on the coefficient $b_3^p$ 
would be $|b_3^p| \lsim 10^{-18}$ eV.
This is more than four orders of magnitude better
than bounds attainable from 1S-2S transitions
and roughly three orders of magnitude better
than estimated attainable bounds 
\cite{bkr}
from $g-2$ experiments in Penning traps.

To summarize,
\it
unsuppressed Zeeman hyperfine spectroscopic signals
for Lorentz or CPT violation appear for transitions
involving spin-flip hyperfine states in hydrogen and antihydrogen
atoms confined in a magnetic trap with an axial bias field
\cite{bkr2}.
\rm

\section*{ACKNOWLEDGMENTS}
 
V.A.K.\ thanks Orfeu Bertolami, Don Colladay,
Rob Potting, Stuart Samuel, and Rick Van Kooten
for collaborations leading to some of the results
described in this talk.
This work is supported in part 
by the Department of Energy
under grant number DE-FG02-91ER40661 
and by the National Science Foundation
under grant number PHY-9503756.

\end{document}